\UseRawInputEncoding
\documentclass[aps,prl,twocolumn,floatfix,groupedaddress]{revtex4-2}
\usepackage{amsmath}
\usepackage{amssymb}
\usepackage{graphicx}
\usepackage{color}
\usepackage[breaklinks=true,colorlinks=true,allcolors=blue]{hyperref}

\begin{document}
\title{The Effective Single-mode model of a Binary Boson Mixture in the Quantum Droplet Region}
\author{Yuncheng Xiong}
\author{Lan Yin}
\affiliation{School of Physics, Peking University, Beijing 100871, China}
\date{\today}

\begin{abstract}
In a binary quantum droplet,  the interspecies attraction dominates over the intraspecies repulsions and the mean-field energy is unstable.  The mechanical stability is restored by the repulsive Lee-Huang-Yang (LHY) energy \cite{petrov2015}.  In the Bogoliubov theory of the binary quantum droplet, there are two branches of gapless excitations.  The lower branch describes the phonon excitation and its energy is imaginary in the long-wavelength limit, implying dynamical instability.  Recently it is found that the phonon energy is renormalized by higher-order quantum fluctuations and the dynamical instability is removed \cite{gu2020}.  In this work, we study a binary quantum droplet in the path-integral formalism to construct an effective model with the correct phonon energy.  By integrating out the upper excitation branch, we obtain an effective single-mode model describing density fluctuations, and derive the extended Gross-Pitaevskii equation.  In this approach, the LHY energy in the extended-GP equation is purely positive without any assumption of neglecting the imaginary part.  This effective single-mode model can be also used outside and close to the quantum droplet region such as in the LHY gas.
\end{abstract}

\maketitle

\paragraph{Introduction.} Quantum droplets of ultracold atoms have been realized in dipolar Bose gases \cite{kadau2016,ferrier2016,chomaz2016,schmitt2016,trautmann2018} and binary boson mixtures \cite{cabrera2018,cheiney2018,semeghini2018}.  The binary quantum droplet is formed in the mean-field unstable region where intraspecies interactions are repulsive, the interspecies interaction is attractive, and the total mean-field energy is attractive.  The repulsive LHY energy balances the attractive mean-field energy and restores the mechanical stability \cite{petrov2015}.  Various theoretical methods have been applied to investigate the properties of the binary quantum droplet, such as the extended Gross-Pitaevskii equation (EGPE) \cite{gautam2019,fort2021,ferioli2020}, quantum Monte Carlo method \cite{cikojevic2018,parisi2019}, and variational theory \cite{staudinger2018,otajonov2019,astrakharchik2018,cappellaro2018}.  Quantum droplets with nontrivial internal structures emerge through the interplay of mean-field energy, Lee-Huang-Yang (LHY) correction and external fields, such as with spin-orbit coupling \cite{liyongyao2017,sachdeva2020,sanchez2020,xiong2021} or confinement \cite{cui2021}.   The self-bound droplet states are also studied in binary dipolar boson mixtures \cite{boudjemaa2018,bisset2021,smith2021a,smith2021b}.

\par The current microscopic theory of the binary quantum droplet is incomplete.  In the Bogoliubov theory, there are two branches of gapless excitations.  The phonon excitation is described by the lower branch, and its energy becomes imaginary in the long-wavelength limit, suggesting dynamical instability.  It was postulated that the dynamical instability can be removed after integrating out high-energy excitations \cite{petrov2015}.  The LHY energy of the droplet calculated in the Bogoliubov theory is complex, but its imaginary part is small and ignored in the construction of the EGPE. Recently it is found that the phonon energy of the binary quantum droplet in the Beliaev theory is real, showing that the dynamical instability is removed by higher-order quantum fluctuations \cite{gu2020}.  There are other proposals that the ground state of the binary quantum droplet is not a Bose-Einstein-condensation (BEC), but a Bardeen-Cooper-Schrieffer (BCS) state \cite{hu2020prl,hu2020pra1,hu2020pra2} or a squeezed coherent state \cite{wangyuqi2020}.

In this work, we study the binary boson mixture at zero-temperature in the path-integral formalism both in and near the quantum droplet region.   The path-integral method has been used to study a single-component Bose gas at zero \cite{braaten1999,andersen2004} and finite temperatures \cite{haugset1998}.  Our goal is to obtain an effective model correctly describing excitations of the lower branch.  For this purpose, we integrate out the excitations of the upper branch in the one-loop order and obtain an effective single-mode action.  By diagonalization we find that the phonon energy is renormalized to a positive value in the droplet region in agreement with Ref. \cite{gu2020}.  Close to the droplet region, the phonon energy is linearly dispersed and well behaved as well.  Particularly, we obtain the phonon speed of LHY gas proportional to $n_0^{3/4}$ where $n_0$ is the condensate density of the Bose gas.  In this approach, the LHY energy which is totally positive.  For the nonuniform case, we derive the EGPE in agreement with Ref. \cite{petrov2015}, without the problem of the imaginary part of the LHY energy.  Our results about the lower-branch excitations can be tested in current experiments.

\paragraph{Path-Integral Formalism} In the path-integral formalism \cite{negele2018}, the grand partition function of a homogeneous binary boson mixture at zero temperature is given by
\begin{equation}
\mathcal{Z}=\int \mathcal{D}[\psi^{\ast},\psi^{}] e^{iS[\psi^\ast,\psi]},
\label{eq:partition}
\end{equation}
where the action is given by
\begin{align}
S[\psi^\ast,\psi]=\int dt d\mathbf{r}\sum\limits_{\sigma=a,b}\psi^\ast_\sigma&(i \partial_t+\frac{\vec{\nabla}^2}{2m_\sigma}+\mu_\sigma \nonumber\\
&-\frac{1}{2}\sum\limits_{\sigma^\prime=a,b}g_{\sigma\sigma^\prime}\psi^\ast_{\sigma^\prime}\psi_{\sigma^\prime})\psi_\sigma,
\label{eq:action_ab}
\end{align}
where $\psi_\sigma$ is the boson field of component $\sigma$, $\mu_\sigma$ is the chemical potential, and $g_{\sigma\sigma'}=g_{\sigma'\sigma}$ is the s-wave coupling constant.  In this work, we consider the case where the two boson components have the same mass, $m_a=m_b=m$, with zero detuning energy, $\mu_a=\mu_b=\mu$.  We assume that the interspecies interaction is attractive, $g_{ab}<0$, and the intraspecies interactions are repulsive, $g_{aa}>0$ and $g_{bb}>0$, according to experiments on $^{39}$K.  For convenience we set $m=1$ and $\hbar=1$.

In the Bose-Einstein-condensation (BEC) state, the boson field $\psi_\sigma$ can be separated into its condensation part $\nu_\sigma$ and fluctuation part $\phi_\sigma$, $\psi_\sigma=\nu_\sigma+\phi_\sigma$.  The condensation part satisfies the minimum condition of the thermodynamic potential, $\partial (E_0-\mu N)/\partial \nu_{\sigma}=0$, where $E_0$ is the ground-state energy and $N$ is the total boson number.  Without losing generality, we take $\nu_\sigma$ as a positive number.    The condensation density of each component is given by $n_{0\sigma}=\nu_\sigma^2$.  Here we consider the dilute case where the ground-state energy and chemical potential as well as other thermodynamic quantities can be expanded in terms of the total condensate density $n_0=n_{0a}+n_{0b}$.  In these expansions, the leading terms are the mean-field ground-state energy, $$E_0^{(0)}=\frac{V}{2}\sum\limits_{\sigma,\sigma^\prime}g_{\sigma\sigma^\prime}n_{0\sigma}n_{0\sigma^\prime},$$
and the mean-field chemical potential $\mu_0$, where $V$ is the volume.  The leading order of the saddle-point condition is given by
\begin{equation}\label{mu0}
\mu_0=\frac{\partial E_0^{(0)}}{V\partial n_{0\sigma}}=g_{\sigma\sigma}n_{0\sigma}+g_{\sigma\bar{\sigma}}n_{0\bar{\sigma}},
\end{equation}
where $\bar{\sigma}$ labels the component different from $\sigma$.
From Eq. (\ref{mu0}), the density ratio of the two condensate components can be determined,
\begin{equation}
\frac{n_{0a}}{n_{0b}}=\frac{g_{bb}-g_{ab}}{g_{aa}-g_{ab}}.
\end{equation}
where at the mean-field unstable point $g_{ab}=-\sqrt{g_{aa}g_{bb}}$, the density ratio is given by $n_{0a}/n_{0b}=\sqrt{g_{bb}/g_{aa}}$ as in Ref. \cite{petrov2015}.

Beyond the mean field, the leading-order fluctuation is the gaussian fluctuation, and in the action is described by
\begin{align}
S^{(2)}&\!=\!\int\! dt d\mathbf{r} \sum\limits_\sigma\bigg\{ \phi^\ast_\sigma\big(i\partial_t+\frac{\vec{\nabla}^2}{2m}+\mu_{0}\big)\phi_\sigma \\
&-\sum\limits_{\sigma'} \frac{g_{\sigma\sigma'}}{2}\nu_\sigma\nu_{\sigma'}\big[2\phi^\ast_\sigma(\frac{\nu_{\sigma'}}{\nu_\sigma}\phi_{\sigma}+\phi_{\sigma'})+(\phi^\ast_\sigma\phi^\ast_{\sigma'}+c.c.)\big]\bigg\},
\nonumber
\end{align}
where the chemical-potential difference $\delta \mu=\mu-\mu_0$ is due to higher-order fluctuations and neglected to this order.  The quadratic Gaussian action can be readily diagonalized, yielding two branches of excitations consistent with the Bogoliubov theory of a binary boson mixture \cite{larsen1963}
\begin{align}
\varepsilon_{\pm}(\mathbf{k})=\Big[\epsilon_{\mathbf{k}}\Big(\epsilon_{\mathbf{k}}+g_{aa}n_{0a}+g_{bb}n_{0b}\pm \lvert g_{ab}n_0\rvert\Big)\Big]^{1/2}
\label{eq:spectrum-ab},
\end{align}
where $\epsilon_{\mathbf{k}}=k^2/2$.  The two excitation branches can be integrated out, generating the LHY energy
\begin{align}
\varepsilon_{\text{LHY}}=&\frac{V}{2}\int\frac{d\mathbf{k}}{(2\pi)^3}\big[\varepsilon_{+}(\mathbf{k})+\varepsilon_{-}(\mathbf{k})-\sum_\sigma (\epsilon_{\mathbf{k}}+g_{\sigma\sigma}n_{0\sigma}) \nonumber\\
&+\frac{1}{2\epsilon_{\mathbf{k}}}(n_{0a}^2 g_{aa}^2+n_{0b}^2 g_{bb}^2+2n_{0a} n_{0b} g_{ab}^2)\big] \nonumber \\
=&\frac{8}{15\pi^2}[\mu_0^{5/2}+(\mu_0+\lvert g_{ab}\rvert n_0)^{5/2}].
\label{eq:LHY}
\end{align}

\par As in the Bogoliubov theory of a binary quantum droplet \cite{petrov2015}, with $\delta g=g_{ab}+\sqrt{g_{aa}g_{bb}}<0$,  the excitation energy in the lower branch is imaginary in the long-wavelength limit, $\varepsilon^2_{-}(\mathbf{k})<0$, implying the dynamical instability.  Hence, the LHY energy given in Eq. (\ref{eq:LHY}) is complex.  In the long-wavelength limit, the excitation energy of the upper branch is much larger than that of the lower branch in magnitude, $\varepsilon_{+}(\mathbf{k}) \gg |\varepsilon_{-}(\mathbf{k})|$.  In Ref. \cite{petrov2015}, Petrov postulated that in the quantum droplet region the excitations in the lower branch could be stabilized by integrating out high-energy excitations, and argued that the imaginary part of the LHY energy can be neglected because it is very small.  Recently in a Beliaev approach, it is found that the excitation energy of the lower branch is renormalized due to interactions with the upper branch and the dynamical instability is removed by higher-order fluctuations \cite{gu2020}.  In the following, we are going to build an effective model of the lower branch and construct the EGPE without the problem of complex LHY energy.

\paragraph{Effective Model of the Lower Branch.}
To separate the quasi-particles in the lower branch from the upper branch, we perform a canonical transformation and define new boson fields given by $\psi_c=(\nu_a \psi_a+\nu_b \psi_b)/\sqrt{n_0}$ and $\psi_d=(\nu_b \psi_a-\nu_a \psi_b)/\sqrt{n_0}$ \cite{gu2020}.  As demonstrated below, the $\psi_c$-field  describes the lower-branch excitations and the $\psi_d$-field describes the upper branch.
In terms of these new fields, the action is given by
\begin{align}
S&=\int dt d\mathbf{r}\bigg[\sum\limits_{i}\psi^\ast_i\bigg((i\partial_t+\frac{\vec{\nabla}^2}{2m})+\mu \bigg)\psi_i \nonumber\\
&-\frac{1}{2}\bigg(g_1\left|\psi^{}_c\right|^4 + g_2\left|\psi^{}_d\right|^4 + g_5\left|\psi^{}_c\right|^2 \left|\psi^{}_d\right|^2 \nonumber\\
&+\big( g_4\psi^{\ast 2}_c \psi^{2}_d +g_6\psi^{\ast}_c \psi^{}_d \left| \psi^{}_d\right|^2+c.c.\big)\bigg)\bigg],
\label{eq:action_cd}
\end{align}
where $i=c,d$, and the expressions of new coupling constants $g_i$ are given in the appendix.  In the BEC phase, $\langle\psi_c\rangle=\sqrt{n_0}$ and $\langle\psi_d\rangle=0$, the action can be further written in terms of the fluctuation fields $\phi_c=\psi_c-\sqrt{n_0}$ and $\phi_d=\psi_d$,
\begin{align}
S=S_0+S_c+S_d+S_{cd},
\end{align}
where the constant part $S_0$ is given by
\begin{align}
S_0&=-\int dt d\mathbf{r}(\frac{E_0^{(0)}}{V}-\mu n_0 ),
\end{align}
the terms $S_c$ and $S_d$ describe only the $\phi_c$- and $\phi_d$-fields respectively, and the term $S_{cd}$ describes the coupling between the two fields,
\begin{widetext}
	\begin{align}
	S_c=\int dt d\mathbf{r} \Big[\phi^\ast_c(i\partial_t+\frac{\nabla^2}{2m}+\mu-2g_1 n_0)\phi_c  -\frac{1}{2}g_1\Big(n_0 (\phi_c\phi_c+c.c.)+2\sqrt{n_0} \phi^\ast_c\phi_c(\phi^\ast_c+\phi_c)+(\phi^\ast_c\phi_c)^2\Big)\Big],
	\label{eq:action_cd_cpart}
	\end{align}
	\begin{align}
	S_d&=\int dt d\mathbf{r} \Big[\phi^\ast_d(i\partial_t+\frac{\vec{\nabla}^2}{2m}+\mu-\frac{1}{2}g_5 n_0)\phi_d-\frac{1}{2}g_4 n_0(\phi_d\phi_d+c.c.)-\frac{1}{2}g_6 \sqrt{n_0}\phi^\ast_d \phi_d(\phi^\ast_d+\phi_d)-\frac{1}{2}g_2 (\phi^\ast_d\phi^\ast_d)^2\Big],
	\end{align}
	\begin{align}
	S_{cd}=-\frac{1}{2}&\int dt d\mathbf{r} \Big[2g_4 \sqrt{n_0}(\phi^\ast_c\phi_d\phi_d+c.c.)+g_5 \sqrt{n_0}(\phi^\ast_c+\phi_c)\phi^\ast_d\phi_d+g_4(\phi^\ast_c\phi^\ast_c\phi_d\phi_d+c.c.) +g_5\phi^\ast_c\phi^\ast_d\phi_d\phi_c+g_6(\phi^\ast_c\phi^\ast_d\phi_d\phi_d+c.c.)\Big].
	\label{eq:action_cd_interaction}
	\end{align}
\end{widetext}
The $\phi_c$- and $\phi_d$-fields are completely decoupled in the guassian fluctuations which can be further diagonalized yielding the lower- and upper-branch excitation spectra to the leading order as in Eq. (\ref{eq:spectrum-ab}).

In and close to the quantum droplet region, the upper-branch excitation is well behaved and can be integrated out to obtain an effective model of the lower branch.  In the partition function,
\begin{align}
\mathcal{Z}=e^{i S_0}\int \mathcal{D}[\phi^\ast_c,\phi_c]e^{i S_c} \int \mathcal{D}[\phi^\ast_d,\phi_d]e^{i(S_d+S_{cd})},
\label{eq:}
\end{align}
the $\phi_d$-field can be integrated out leading to
\begin{align}
\mathcal{Z}=e^{i(S_0+\delta S_0)}\int \mathcal{D}[\phi^\ast_c,\phi_c]e^{i(S_c+\delta S_c)},
\label{eq:effective}
\end{align}
where $\delta S_0$ and $\delta S_c$ are the corrections in the action.  For a dilute gas, it is sufficient to keep only the guassian terms in $S_{d}$.  Thus the propagator of the $\phi_d$-field in the momentum space is given by
\begin{equation}
\begin{split}
G_{d^\ast d}(k)&=\frac{k_0+\epsilon_{\mathbf{k}}+g_4 n}{k_0^2-\varepsilon_+^2(\mathbf{k})+i\eta}, \\
G_{d^\ast d^\ast}(k)&=G_{dd}(k)=\frac{-g_4 n}{k_0^2-\varepsilon_+^2(\mathbf{k})+i\eta},
\label{eq:green}
\end{split}
\end{equation}
where $k=(k_0,\mathbf{k})$, $k_0$ is the frequency, $G_{d^\ast d}(k)$ is normal (diagonal) part of the propagator, and $G_{d^\ast d^\ast}(k)=G_{dd}(k)$ is the anomalous (off-diagonal) part.  In the procedure of integrating out the $\phi_d$-field, it is sufficient to consider only the one-loop order.  Thus we obtain the correction to the constant part of the action corresponding to the upper-branch part of the LHY energy given by
\begin{equation}
E_d=\frac{8V}{15\pi^2}(g_4 n_0)^{5/2},
\label{U-LHY}
\end{equation}
as in Eq. (\ref{eq:LHY}) from the Bogoliubov theory, and the correction to the mean-field chemical potential given by
\begin{equation}
\mu_d=\frac{32}{3\sqrt{\pi}}\sqrt{n_0 a_4^3}g_4 n_0.
\end{equation}
where $g_4\equiv4\pi a_4$.

In the effective action of the $\phi_c$-field, $S_c^{\text{(eff)}}=S_c+\delta S_c$, the terms beyond the quadratic order are unimportant and can be neglected in the dilute limit,
\begin{align}
S_c^{\text{(eff)}} &\approx \int \frac{d k_0}{2\pi}\sum_{\bf k} \Big[ \phi^\ast_c(k)\Big(k_0-\epsilon_{\mathbf{k}}+\mu_c-\Sigma_{11} \Big)\phi_c(k) \nonumber\\
&-\frac{1}{2}\Sigma_{20}\phi^\ast_c(k)\phi^\ast_c(-k)-\frac{1}{2}\Sigma_{02}\phi_c(k)\phi_c(-k)\Big],
\label{eq:action_effective}
\end{align}
where $\Sigma_{11}$ is the normal part of the self-energy, $\Sigma_{20}$ ($\Sigma_{02}$) is the anomalous part, and $\mu_c=\mu_0+\mu_d$ is the corrected chemical potential.
\begin{figure}[hbt]
	\includegraphics[width=0.9\columnwidth]{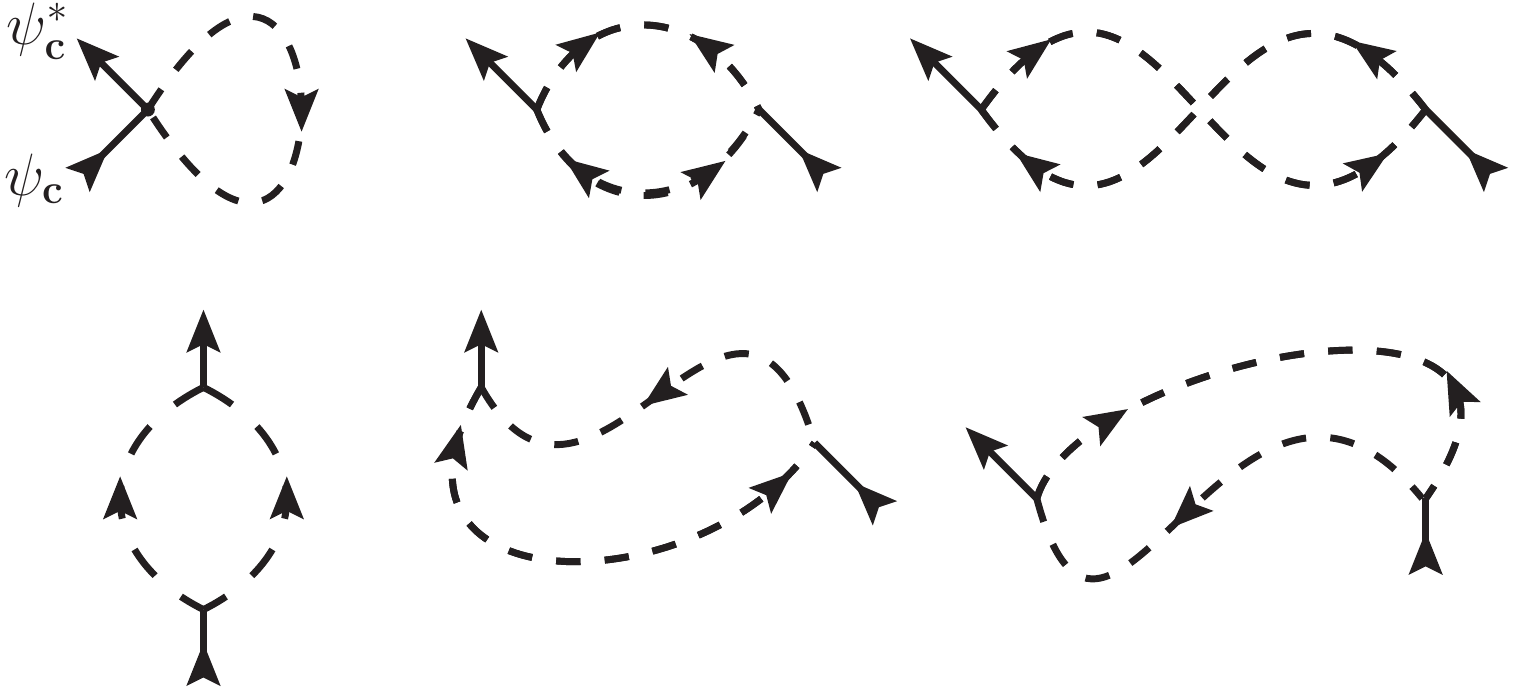}
	\caption{The one-loop diagrams of $\Sigma_{11}$.  Dashed lines with one-way arrow(s) denote the normal Green's function $G_{d^\ast d}$ while those with two head-to-head (tail-to-tail) arrows denote the anomalous  the Green's function $G_{d^\ast d^\ast}(G_{dd})$.   The external uncontracted solid half-lines denote the boson fields $\psi_c^\ast$ (outgoing) and $\psi_c$ (incoming).}
	\label{fig:sigma11}
\end{figure}
The one-loop Feynman diagrams contributing to $\Sigma_{11}$ and $\Sigma_{20}$ are presented in Fig.\ref{fig:sigma11} and Fig.\ref{fig:sigma20} respectively. The diagrams of $\Sigma_{02}$ (not shown) can be easily drawn by reversing the directions of all arrows in the diagrams of $\Sigma_{20}$.  The expressions of these self-energy diagrams are given by
\begin{figure}
	\includegraphics[width=0.9\columnwidth]{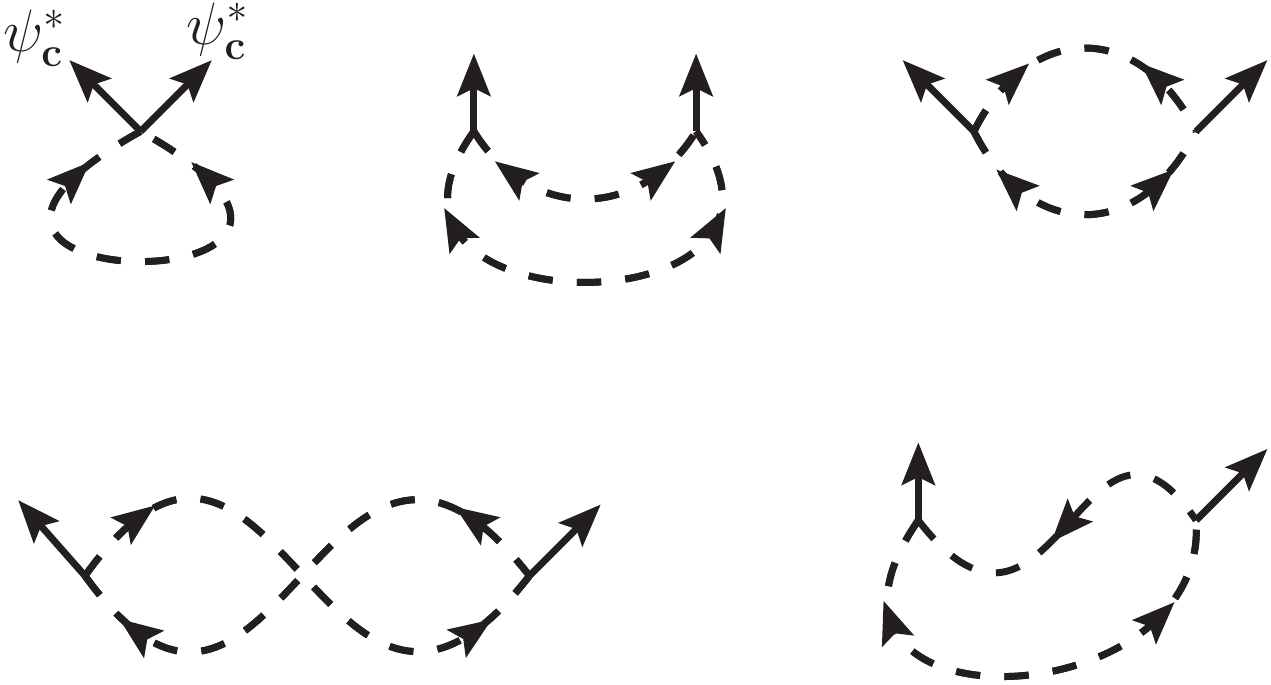}
	\caption{The one-loop diagrams of $\Sigma_{20}$. The denotation is same as in Fig.\ref{fig:sigma11}.}
	\label{fig:sigma20}
\end{figure}
\begin{align}
&\Sigma_{11}(k)=2(g_1+\delta g_1)n_0 +i\int\frac{dq^4}{(2\pi)^4}
\bigg\{\frac{1}{2}g_5 G_{d^\ast d}(q) \nonumber\\
&+\frac{1}{4}g_5^2 n_0\Big[G_{dd}(k+q)G_{d^\ast d^\ast}(q)+G_{d^\ast d}(k+q)G_{d^\ast d}(q)\Big] \nonumber\\
&+2g_4^2 n_0 G_{d^\ast d}(-q)G_{d^\ast d}(k+q) +g_4 g_5 n_0 G_{d^\ast d}(k+q)G_{dd}(q) \nonumber\\
&+g_4 g_5 n_0 G_{d^\ast d}(k+q)G_{d^\ast d^\ast}(q)\bigg\},
\label{eq:sigma11}
\end{align}
\begin{align}
&\Sigma_{20}(k)=(g_1+\delta g_1)n_0 +i\int\frac{dq^4}{(2\pi)^4} \bigg\{g_4 G_{d^\ast d}(q) \\
&+2g_4^2 n_0 G_{d^\ast d}(q)G_{d^\ast d}(k+q) +\frac{1}{4}g_5^2 n_0\Big(G_{dd}(k+q)G_{d^\ast d^\ast}(q)\nonumber\\
&+G_{d^\ast d}(k+q)G_{d^\ast d}(q)\Big)+2g_4 g_5 n_0 G_{d^\ast d}(-q) G_{dd}(k+q) \bigg\},  \nonumber
\label{eq:sigma20}
\end{align}
where the terms with $g_4$ or $g_5$ are from the one-loop diagrams, the terms with $g_1$ are tree-level contributions, and the term with
$$\delta g_1=(2g_4^2-g_4 g_5+\frac{1}{4}g_5^2)\int\frac{d\mathbf{k}}{(2\pi)^3}\frac{1}{2\epsilon_{\mathbf{k}}}$$
is due to renormalization of coupling constants.

In the low-energy and long-wavelength limit, the self-energy is given by
\begin{align}
\Sigma_{11}(0)&=2g_1 n_0+\frac{80}{3\sqrt{\pi}}\sqrt{n_0 a_4^3}g_4 n_0, \nonumber \\
\Sigma_{20}(0)&=g_1 n_0+\frac{16}{\sqrt{\pi}}\sqrt{n_0 a_4^3}g_4 n_0,
\end{align}
which satisfies the Pines-Hugenholtz theorem  $$\mu_c=\mu_0+\mu_d=\Sigma_{11}(0)-\Sigma_{20}(0).$$
Thus the low-energy excitations of the lower branch can be described by the effective action,
\begin{align}
S_c^{(\text{eff})}& \approx \int \frac{d k_0}{2\pi}\sum_{\bf k} \Big[ \phi^\ast_c(k)\Big(k_0-\frac{\mathbf{k}^2}{2}-\Sigma_{20}(0)
\Big)\phi_c(k) \nonumber\\
&-\frac{1}{2}\Sigma_{20}(0)\phi^\ast_c(k)\phi^\ast_c(-k)-\frac{1}{2}\Sigma_{20}(0)\phi_c(k)\phi_c(-k)\Big],
\label{eq:action_effective2}
\end{align}
which is consistent with the Bogoliubov-de Gennes equation (BDGE) shown below.

\paragraph{Phonon energy and LHY energy.}  From the effective action $S_c^{(\text{eff})}$ in Eq. (\ref{eq:action_effective2}), we obtain the renormalized excitation energy of the lower branch in the long-wavelength limit given by
\begin{align}
k_0=\sqrt{\frac{\mathbf{k}^2}{2}\big(\frac{\mathbf{k}^2}{2}+2g_1 n_0+\frac{32}{\sqrt{\pi}}\sqrt{n_0 a_4^3}g_4 n_0\big)}, \label{eq:new_phonon_spectrum_final}
\end{align}
which is applicable in and close to the droplet region as long as the lower branch and the upper branch are well separated at low energies. In the droplet region, the condensate density to the leading order of $\delta g$ is given by \cite{petrov2015}
$$n_0=\frac{25\pi}{1024}\frac{\lvert\delta g\rvert^2}{(a_{aa}a_{bb})^{3/2}(\sqrt{g_{aa}}+\sqrt{g_{bb}})^4},$$
and the phonon energy can be further rewritten as
\begin{equation}
k_0=\sqrt{\frac{\mathbf{k}^2}{2}\big(\frac{\mathbf{k}^2}{2}+\frac{32}{5\sqrt{\pi}}\sqrt{n_0 a_4^3}g_4 n_0\big)}.
\end{equation}
The sound velocity, i.e. the phonon speed, is real and positive, given by
\begin{equation}
c_{d}=\sqrt{\frac{16}{5\sqrt{\pi}}\sqrt{n_0 a_4^3}g_4 n_0}, \label{cd}
\end{equation}
as in Ref. \cite{gu2020}, showing the system is dynamically stable.
For a LHY gas with $\delta g=g_1=0$, $g_4=\sqrt{g_{aa}g_{bb}}$, the phonon energy given by Eq.(\ref{eq:new_phonon_spectrum_final}) can be simplified as
\begin{equation}
k_0=\sqrt{\frac{\mathbf{k}^2}{2}\big(\frac{\mathbf{k}^2}{2}+\frac{32}{\sqrt{\pi}}\sqrt{n_0 a_4^3}g_4 n_0\big)},
\end{equation}
and the sound velocity is given by
\begin{equation}\label{cg}
c_{g}=\sqrt{\frac{16}{\sqrt{\pi}}\sqrt{n_0 a_4^3}g_4 n_0}.
\end{equation}
Although the sound velocity is proportional to $n_0^{3/4}$ in both cases, there is a jump in sound velocity when the system goes from the droplet state to LHY gas with $\delta g$ approaching to zero from the negative side.  The sound velocity is the Landau critical velocity \cite{lifshitz1980} which have been measured in a single-component BEC \cite{andrews1997,raman1999}.  The same experimental technique can be used in current experiments on $^{39}K$ to measure the sound velocity of the quantum droplet.

The LHY energy of the lower branch can be obtained by integrating out the $\phi_c$-field in Eq. (\ref{eq:action_effective2}), given by
\begin{equation}
E_c=\frac{8V}{15\pi^2}\big(g_1 n_0 + \frac{16}{\sqrt{\pi}}\sqrt{n_0 a_4^3}g_4 n_0\big)^{5/2}.
\label{L-LHY}
\end{equation}
In the droplet region, contrary to the Bogoliubov theory \cite{petrov2015}, the LHY energy of the lower branch in our approach is purely real without any imaginary part.  In and close to the droplet region, the lower-branch part of the LHY energy given in Eq. (\ref{L-LHY}) is  much smaller than the upper-branch part in Eq. (\ref{U-LHY}), and can be safely ignored as argued in Ref.  \cite{petrov2015}.  It is worth to note that for a dilute droplet, $n_0a_4^3\ll1$, integrating out the $\phi_d$-field also generates two-loop correction to the ground state energy proportional to $Vn_0^3a_4^4$, much bigger than the LHY energy of the lower branch which is proportional to $Vn_0^{15/4}a_4^{25/4}$.  Thus the important next-order fluctuations come from the two-loop diagrams of the $\phi_d$-propagator, instead of the one-loop diagrams of the $\phi_c$-propagator.

\paragraph{Generalization to the nonuniform case.}
Gross-Pitaevskii equation (GPE) has been widely used to simulate weakly interacting Bose gases \cite{pethick2008,pitaevskii2016}.  In GPE, the interaction is treated in the mean-field level. However in the droplet regime, the real part of the LHY energy in the Bogoliubov theory is comparable with the mean-field energy and has been simply put into GPE with the imaginary part of the LHY energy neglected \cite{petrov2015}, leading to the EGPE.  In our approach, the LHY energy is real, and the EGPE can be derived without the issue of complex LHY energy.

The action of an inhomogeneous binary Bose gas in a time-dependent trap $V({\bf r},t)$ is given by
\begin{align}
S_I=&\int dt d\mathbf{r}\sum\limits_{\sigma}\psi^\ast_\sigma(i \partial_t+\frac{\vec{\nabla}^2}{2m_\sigma}-V({\bf r},t) \nonumber\\
&+\mu-\frac{1}{2}\sum\limits_{\sigma^\prime}g_{\sigma\sigma^\prime}\psi^\ast_{\sigma^\prime}\psi_{\sigma^\prime})\psi_\sigma.
\label{eq:action_Iab}
\end{align}
Here we consider the case that the system is slowly varying in space and time so that local density approximation (LDA) can be applied and local equilibrium can be assumed.  In LDA, the real space can be divided into many small subspaces and each subsystem can be treated as a uniform system.  With the local equilibrium assumption, each subsystem is assumed in its ground state locally.  The fluctuations in the boson fields around the the condensate wavefunction  $\langle \psi_\sigma({\bf r},t) \rangle =\nu_\sigma({\bf r},t)$ can be integrated out locally as prescribed in the above sections, leading to the effective action of the single-mode condensate,
\begin{align}
S_I^{\text{(eff)}}&=\int dt d\mathbf{r}\big[\sum\limits_{\sigma}\nu^\ast_\sigma(i \partial_t+\frac{\vec{\nabla}^2}{2m_\sigma}-V({\bf r},t) \nonumber\\
&+\mu-\frac{1}{2}\sum\limits_{\sigma^\prime}g_{\sigma\sigma^\prime}\nu^\ast_{\sigma^\prime}\nu_{\sigma^\prime})\nu_\sigma-\frac{E_d(n_0)}{V}\big],
\end{align}
where the last term is the LHY energy of the local upper branch given by Eq. (\ref{U-LHY}), $n_0({\bf r},t)=|\nu_a({\bf r},t)|^2+|\nu_b({\bf r},t)|^2$ is the condensate density, and the LHY energy of the local lower branch is neglected due to its insignificance in and close to the quantum droplet region.
Here we restrict to the single-mode case, $\nu_\sigma({\bf r},t)= e^{i\theta_\sigma}
\sqrt{\eta_\sigma} \Phi_c({\bf r},t)$,
where $\Phi_c({\bf r},t)$ is the single-mode wavefunction of the condensate, $\theta_\sigma$ is the constant phase, and $\eta_\sigma$ is the fraction of the $\sigma$-component condensate, 
$$\eta_\sigma=\frac{g_{\bar{\sigma}\bar{\sigma}}-g_{\sigma\bar{\sigma}}}{g_{\sigma\sigma}+g_{\bar{\sigma}\bar{\sigma}}-2g_{\sigma\bar{\sigma}}}.$$
The single-mode condensate wavefunction is subject to equation of motion,
\begin{align}
\frac{\delta S_I^{\text{(eff)}}}{\delta \Phi_c^\ast}&=\big(i \partial_t+\frac{\vec{\nabla}^2}{2m_\sigma}-V({\bf r},t) \nonumber\\
&+\mu-g_1 |\Phi_c|^2-\frac{d (E_d/V)}{dn_0} \big)\Phi_c({\bf r},t)=0,
\end{align}
which can be further written as
\begin{align}
i\frac{\partial}{\partial t}\Phi_c=(-\frac{\nabla^2}{2}
+V({\bf r},t)+g_1 \lvert\Phi_c\rvert^2+\frac{4}{3\pi^2}g_4^{5/2}\lvert\Phi_c\rvert^3)\Phi_c,
\label{eq:eGPE}
\end{align}
in agreement with Ref.  \cite{petrov2015}.
The excitations around the ground state can be solved from the EGPE as well by assuming $\Phi_c({\bf r})=\Phi_c^{(0)}({\bf r})+\phi_c({\bf r})$ and linearizing both sides of the EGPE with $\phi_c({\bf r})$,
\begin{align} \label{BDGE}
i\frac{\partial}{\partial t}\phi_c=&(-\frac{\nabla^2}{2}\phi_c+V({\bf r},t)
+2g_1 \lvert\Phi_c^{(0)}\rvert^2+\frac{10}{3\pi^2}g_4^{5/2} \lvert\Phi_c^{(0)}\rvert^3)\phi_c \nonumber\\
&+(g_1 \lvert\Phi_c^{(0)}\rvert^2+\frac{2}{\pi^2}g_4^{5/2} \lvert\Phi_c^{(0)}\rvert^3)\phi_c^\ast,
\end{align}
where $\Phi_c^{(0)}({\bf r})$ is the wavefunction of the ground state.  In the stationary case, Eq. (\ref{BDGE}) becomes the extended Bogoliubov-de Gennes equation, and in the uniform case it recovers the extreme condition of the effective action given by Eq. (\ref{eq:action_effective2}).

\paragraph{Conclusion.} In conclusion, we study a binary boson mixture in and close to the quantum droplet region in the path-integral formalism, and obtain an effective action describing the lower-branch excitations by integrating out the upper-branch.  We obtain a positive phonon speed for a quantum droplet consistent with the result in the Beliaev approach \cite{gu2020}.  For a LHY gas, we obtain a phonon speed proportional to $n_0^{3/4}$.  These results can be tested in current experiments.  In our approach, the LHY energy of the lower branch is purely positive without any imaginary part, and its magnitude is much smaller than that of the upper branch as pointed out in Ref. \cite{petrov2015}.  The EGPE can be constructed without any problem of complex LHY energy.

\appendix*
\section{Relations between coupling constants in a-b and c-d representations}
\par  Following the canonical transformation, the coupling constants in the c-d representation are given by:
\begin{align}
g_1&=g_{aa}\frac{\nu_a^4}{\nu^4}+g_{bb}\frac{\nu_b^4}{\nu^4}+2g_{ab}\frac{\nu_a^2 \nu_b^2}{\nu^4}, \label{app_g1}\\
g_2&=g_{aa}\frac{\nu_b^4}{\nu^4}+g_{bb}\frac{\nu_a^4}{\nu^4}+2g_{ab}\frac{\nu_a^2 \nu_b^2}{\nu^4},\\
g_3&=2(g_{aa}\frac{\nu_a^2}{\nu^2}-g_{bb}\frac{\nu_b^2}{\nu^2}+g_{ab}\frac{\nu_b^2-\nu_a^2}{\nu^2})\frac{\nu_a \nu_b}{\nu^2}, \\
g_4&=(g_{aa}+g_{bb}-2g_{ab})\frac{\nu_a^2 \nu_b^2}{\nu^4}, \\
g_5&=4(g_{aa}+g_{bb})\frac{\nu_a^2 \nu_b^2}{\nu^4}+2g_{ab}\frac{(\nu_a^2-\nu_b^2)^2}{\nu^4}, \\
g_6&=2(g_{aa}\frac{\nu_b^2}{\nu^2}-g_{bb}\frac{\nu_a^2}{\nu^2}+g_{ab}\frac{\nu_a^2-\nu_b^2}{\nu^2})\frac{\nu_a \nu_b}{\nu^2}.
\label{}
\end{align}
With the condition $$\frac{\nu_a^2}{\nu_b^2}=\frac{n_{a0}}{n_{b0}}=\frac{g_{bb}-g_{ab}}{g_{aa}-g_{ab}},$$
these relations can be simplified as
\begin{align}
g_1&=\frac{g_{aa}g_{bb}-g_{ab}^2}{g_{aa}-2g_{ab}+g_{bb}}, \\
g_2&=\frac{g_{aa}^2+g_{bb}^2-g_{aa}g_{bb}-g_{ab}^2}{g_{aa}-2g_{ab}+g_{bb}},  \\
g_3&=0, \\
g_4&=\frac{(g_{aa}-g_{ab})(g_{bb}-g_{ab})}{g_{aa}-2g_{ab}+g_{bb}}, \\
g_5&=\frac{2(2g_{aa}g_{bb}-g_{aa}g_{ab}-g_{bb}g_{ab})}{g_{aa}-2g_{ab}+g_{bb}}, \\
g_6&=\frac{2(g_{aa}-g_{bb})\sqrt{(g_{aa}-g_{ab})(g_{bb}-g_{ab})}}{g_{aa}-2g_{ab}+g_{bb}},
\end{align}
Note that $g_3=0$, leading to the decoupling of two excitation branches in the gaussian level, and these coupling constants are not all independent, for example, $g_5=2(g_1+g_4)$.
\bibliography{effective}

\end{document}